\documentclass[aip,numerical,pop,reprint,amsmath,amssymb,showpacs,letterpaper]{revtex4-1}

\usepackage{graphicx}
\usepackage{dcolumn}
\usepackage{bm}
\usepackage{hyperref}

\newcommand{\vek}[1]{\boldsymbol{#1}}
\newcommand{\uvek}[1]{\hat{\boldsymbol{#1}}}

\newcommand{\deltab}{\delta \vek{B}_\perp}
\newcommand{\bnorm}{\uvek{b}}
\newcommand{\bnormp}{\tilde{\vek{b}}}

\newcommand{\dpar}{\uvek{b} \cdot \nabla}
\newcommand{\dparp}{\bnormp \cdot \nabla}
\newcommand{\bcrossnabla}{\left( \vek{B} \times \nabla B \right) \cdot \nabla}

\begin{document}

\begin{flushleft}
This article has been submitted to Physics of Plasmas. After publication, it can be found at \href{http://pop.aip.org}{http://pop.aip.org} 
\end{flushleft}

\title{Verification of long wavelength electromagnetic modes with a gyrokinetic-fluid hybrid model in the XGC code}

\author{Robert Hager}
 \email{rhager@pppl.gov}
\affiliation{%
Princeton Plasma Physics Laboratory\\
P.O. Box 451, Princeton, NJ 08543, USA
}%
\author{Jianying Lang}
\affiliation{%
Intel Corporation\\
2200 Mission College Blvd. Santa Clara, CA 95054, USA
}
\author{C.S. Chang}%
\affiliation{%
Princeton Plasma Physics Laboratory\\
P.O. Box 451, Princeton, NJ 08543, USA
}
\author{S. Ku}%
\affiliation{%
Princeton Plasma Physics Laboratory\\
P.O. Box 451, Princeton, NJ 08543, USA
}
\author{Y. Chen}%
\affiliation{%
University of Colorado\\
2000 Colorado Avenue, Boulder, CO 80309, USA
}
\author{S. E. Parker}%
\affiliation{%
University of Colorado\\
2000 Colorado Avenue, Boulder, CO 80309, USA
}
\author{M. F. Adams}%
\affiliation{%
Lawrence Berkeley National Laboratory\\
1 Cyclotron Road, Berkeley, CA 94720, USA
}

\date{\today}


\begin{abstract}
As an alternative option to kinetic electrons, the gyrokinetic total-f particle-in-cell (PIC) code XGC1 has been extended to the MHD/fluid type electromagnetic regime by combining gyrokinetic PIC ions with massless drift-fluid electrons analogous to Chen and Parker, Physics of Plasmas \textbf{8}, 441 (2001).
Two representative long wavelength modes, shear Alfv\'{e}n waves and resistive tearing modes, are verified in cylindrical and toroidal magnetic field geometries.
\end{abstract}


\maketitle

%
%

%
This article describes the verification of two important MHD/fluid type, long-wavelength, electromagnetic modes after the addition of an optional kinetic-fluid hybrid model to the gyrokinetic particle-in-cell code XGC1 \cite{sku_2009}.
This work complements -- as a more economical alternative -- the fully implicit, fully kinetic electromagnetic formulation, that is also being developed for XGC1 \cite{ku_aps_2016}.

The importance of MHD/fluid type electromagnetic modes in magnetically confined fusion devices, which operate regularly at moderate to high $\beta = 2\mu_0 P/B^2$ (the ratio of thermodynamic to magnetic pressure), is widely recognized. 
Examples are neoclassical tearing modes \cite{connor_1988}, sawtooth oscillations \cite{von_goeler_1974}, and edge localized modes \cite{huysmans_2005}.
Gyrokinetic electromagnetic codes such as GYRO \cite{gyro,candy_2005}, GS2 \cite{kotschenreuther_1995}, GENE \cite{pueschel_2008}, GEM \cite{parker_2000,chen_2001,chen_2003} have been available with increasing physics capability for more than a decade and have also been used to study those modes.
However, their application to long wavelength MHD/fluid type instabilities has been difficult, especially for PIC codes, due to the so called ``cancellation problem'' \cite{cummings_phd_1995,mishchenko_2004}.
%
Recently, several methods to overcome the cancellation problem with kinetic electrons have been developed for particle codes: the control variate method \cite{hatzky_2007}, a special splitting of the vector potential \cite{mishchenko_2014_1,mishchenko_2014_2} (used e.g. by the EUTERPE code), and split-weight methods \cite{wwlee_2001,chen_2003} (used in GEM, and being further developed in GTS \cite{startsev_aps_2016}).
The XGC1 code \cite{sku_2009} recently demonstrated fully kinetic electromagnetic capability without cancellation problem \cite{ku_aps_2016} using a fully implicit electromagnetic scheme based on the work by Chen and Chac\'{o}n \cite{gchen_2014,gchen_2015,chacon_2016}.
These methods are computationally expensive for long wave length MHD/fluid type modes even without the cancellation problem.
The cheapest way to study these modes is to use fluid electrons instead of electron particles.

Long wavelength electromagnetic physics in the global edge region have so far been studied with fluid and MHD codes (some of them with ad-hoc kinetic ion effect) such as BOUT++ \cite{dudson_2009,liu_2014}, M3D \cite{fu_2006,breslau_2006}, M3D-C1 \cite{ferraro_2009,ren_2015}, and JOREK \cite{czarny_2008,huysmans_2010}, which neglect important effects that drive the plasma to a non-thermal equilibrium.
Since kinetic ion effects on fluid/MHD modes as well as microturbulence are expected to be important in the plasma edge region, e.g. for the physics of edge localized modes (ELMs), kinetic ballooning modes (KBM) and others, we will improve the fluid and MHD approach by coupling gyrokinetic ions to the massless electron fluid hybrid model utilized in the GEM code \cite{parker_2000,chen_2001}.
Although the fluid treatment of the electrons drops some important effects such as the trapped electron mode (TEM), it is still attractive because its implementation is rather straightforward without the cancellation issue, low $k_\perp$ fluid/MHD modes are important for ELM activity, and it is economical with computing time.


%
%
The fluid-kinetic hybrid version of the XGC1 code used for this report combines gyrokinetic ions in the $\delta$f formalism \cite{sku_2009} (which, if done correctly, can be made identical to the total-f formalism \cite{sku_2016}) with massless drift-fluid electrons \cite{chen_2001,chen_2015}.
The electron density continuity equation is given by:
\begin{align}
  \frac{\partial \delta n_e}{\partial t} &=
             -n_0 \left( \vek{B} + \deltab \right)
                \cdot \nabla \left( \frac{\nabla_\perp^2 A_\parallel}{e \mu_0 n_0 B}
                   +\frac{u_{\parallel,i}}{B} \right) \notag \\
            &+ \deltab \cdot \nabla \frac{j_0}{e B} 
                - \vek{v}_E \cdot \nabla \left( n_0 + \delta n_e \right) \notag\\
            &- \frac{2 n_0}{B^3} \bcrossnabla \phi \notag\\
            &+ \frac{2}{e B^3}  \bcrossnabla \delta P_e,
            \label{eq:continuity_eq}
\end{align}
where $\boldsymbol B$ is the axisymmetric background magnetic field, $\deltab = \nabla A_\parallel \times \bnorm$ is the perturbed magnetic field, $\bnorm = \boldsymbol{B}/B$, and $A_\parallel$ is the component of the perturbed vector potential along the background magnetic field; $\mu_0$ is the vacuum permeability; $u_{\parallel,i} = \int \mathrm{d}^3v v_\parallel \delta f_i/n_0$ is the parallel ion fluid flow; $\delta f_i$ is the perturbed ion guiding center distribution function; $n_e = n_0 + \delta n_e$ is the electron density; $j_0 = \bnorm \cdot \nabla \times (\vek{B}/\mu_0)$ is the equilibrium current density; $\delta P_e = e \delta n_e T_{0,e}$ is the perturbed iso-thermal electron pressure, and $\vek{v}_E = \frac{1}{B} \bnorm \times \nabla \phi$ is the $\vek{E} \times \vek{B}$ drift.
We also used the relation $u_{\parallel,e} = (\nabla_\perp^2 A_\parallel) / (e \mu_0 n_0) + u_{\parallel,i}$.
The time evolution of the perturbed vector potential is given by the definition of the electric field and Ohm's law,
\begin{align}
 \frac{\partial A_\parallel}{\partial t} &= -\dpar \phi - E_\parallel, \label{eq:apar_eq} \\
  E_\parallel &= -\frac{\dparp}{e n_0} \delta P_e - \frac{\deltab}{e n_0 B}
                  \cdot \nabla \left( P_{0,e} - e n_0 \phi \right) 
                 +\eta_e j_\parallel. \label{eq:ohms_law}
\end{align}
Here, $\bnormp = \bnorm + \deltab/B$, $P_{0,e} = e n_0 T_{0,e}$ is the background electron pressure, and $j_\parallel = e n_0 (u_{\parallel,i}-u_{\parallel,e})$.
Finally, the gyrokinetic Poisson equation in the long wave length limit is
\begin{equation}\label{eq:poisson_eq}
 -\nabla \cdot \frac{\epsilon_0 \chi}{e} \nabla \phi = \delta n_i - \delta n_e,
\end{equation}
where $\chi=(\rho_i/\lambda_{D,i})^2 = (c/v_A)^2$ is the ion electric susceptibility, $\rho_i$ is the ion gyro radius, $\lambda_{D,i}$ is the ion Debye length, and $v_A = B/(\mu_0 m_i n_0)^{1/2}$ is the Alfv\'{e}n speed.
The ion density is $\delta n_i = \int \mathrm{d}^3v \langle \delta f_i \rangle_\rho$, where $\langle \dots \rangle_\rho$ indicates gyro-averaging.

The massless electron approximation is valid in the limit $v_A/v_{t,e} \rightarrow 0$ or $\beta_e m_i/m_e \gg 1$, where $\beta_e = 2 \mu_0 P_e/B^2$.

Both explicit and implicit time integrators have been implemented.
A second order Runge-Kutta (RK2) method has been utilized for the time integration of the combined particle-fluid system for many of the results discussed in this work.
In the first step, $\delta n_e(t+\Delta t/2)$ and $A_\parallel (t+\Delta t/2)$ are evaluated using, $\phi(t)$, $\delta n_i(t)$ and $u_{\parallel,i}(t)$.
Then the particles are pushed for a half time step to evaluate $\delta n_i(t+\Delta t/2)$ and $u_{\parallel,i}(t+\Delta t/2)$.
In the second step, we evaluate $\phi(t+\Delta t/2)$ and then push the particles for a full time step to obtain $\phi(t+\Delta t)$, $\delta n_i(t+\Delta t)$ and $u_{\parallel,i}(t+\Delta t)$.

Implicit time stepping methods have been implemented using the PETSc TS framework \cite{petsc-web-page,petsc-user-ref, petsc-efficient} to overcome the restrictions on the time step of explicit methods.
The particle terms $\delta n_i$ and $u_{\parallel,i}$ are treated as non-linear contributions to the system of electron fluid equations and are fully integrated into PETSc's nonlinear solver residual, but only the electron fluid terms are included in the Jacobian.
The Newton method is used to solve the non-linear equations, which requires one particle push per evaluation of the residual.

%
%

%
For the verification of shear Alfv\'{e}n wave physics, we use a minimal system that supports this mode: linearized equations versions of \eqref{eq:continuity_eq}-\eqref{eq:ohms_law} with the closure $E_\parallel = \eta_e j_\parallel$.
In addition, we neglect the terms related to the curvature and $\nabla B$ drift in Eq. \eqref{eq:continuity_eq}.
It is straightforward to prove in cylindrical geometry that the dispersion relation of the resulting reduced system yields $\omega = [v_A2 k_\parallel^2 - 4(\eta_e/(2 \mu_0))^4 k_\perp^4 ]^{1/2} + i \eta_e/(2\mu_0) k_\perp^2$.
The first verification test of the shear Alfv\'{e}n dispersion relation was conducted in cylindrical geometry with concentric, circular flux-surfaces with minor radius $a=1$, constant safety factor $q=3$, $\beta_e = 1.5 \cdot 10^{-2}$, and $\eta_e = 10^{-6}\,\Omega\mathrm{m}$.
The simulation was initialized with a global perturbation of $A_\parallel$ centered around $r/a=0.67$ containing toroidal mode numbers $n=1\dots4$ and poloidal mode numbers $m=0\dots4$.
With this large scale variation in the radial and poloidal direction, the low resistivity does not influence the real frequency much but still serves as a check for the resistive dissipation of the reduced shear Alfv\'{e}n wave system (with $k_\perp \sim 1/a$).
The time step for this simulation was $\Delta t = 1.36 \cdot 10^{-8}\,\mathrm{s} \approx 10^{-2} \tau_A$, where $\tau_A = R_0/v_A$.
The total duration of the simulation is $1.36 \cdot 10^{-3}\,\mathrm{s} \approx  1000 \tau_A$.

\begin{figure}
 \centering
 \includegraphics[bb=0 0 228.518097 161.270447]{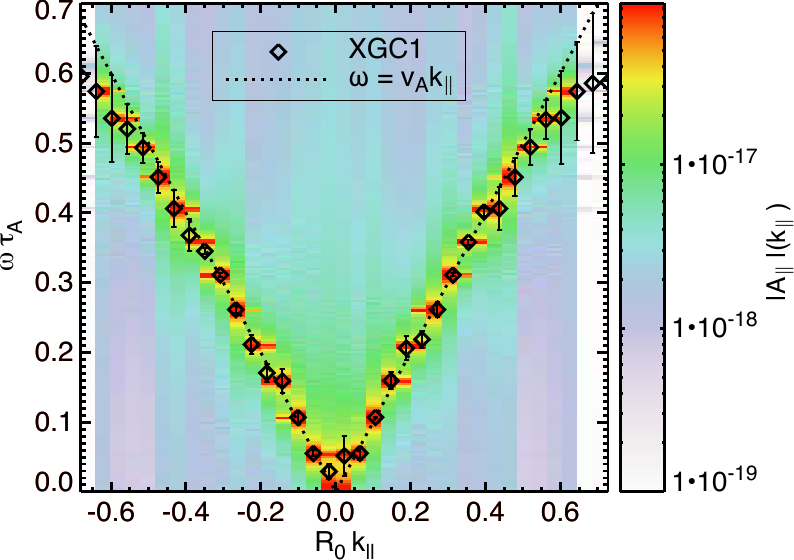}
 \caption{Shear Alfv\'{e}n wave spectrum in cylindrical geometry with concentric circular flux-surfaces. The density plot indicates the mode intensity, the diamonds the median of the intensity at each $k_\parallel$, and the error bars the decay length of the intensity around the median. The steps in the median frequency are an artifact from the interpolation from $(k_\theta,k_\varphi)$ space to a common $k_\parallel$ scale.}
 \label{fig:alfven_dispersion_cylinder}
\end{figure}

Figure \ref{fig:alfven_dispersion_cylinder} shows the shear Alfv\'{e}n spectrum obtained from this simulation.
The parallel wave number was determined as $k_\parallel = \bnorm \cdot \vek{k} = (B_P/B) k_\theta + (B_T/B) k_\varphi$.
The mode frequency is the median of the intensity for each value of $k_\parallel$ and the error bars indicate the decay length of the mode intensity around its median.
The increasing width of the error bars at $k_\parallel > 0.5$ indicates decreasing overall intensity due to the low toroidal and poloidal mode numbers used to initialize the simulation.
The steps in the frequency spectrum are an artifact of the interpolation of the intensity from $(k_\theta,k_\varphi)$ space to a common $k_\parallel$ scale.

\begin{figure}
 \centering
 \includegraphics[bb=0 0 202.004959 146.555405]{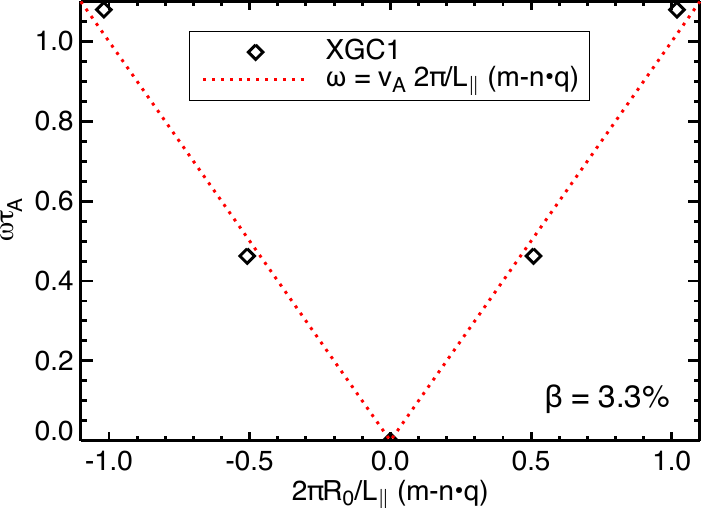}
 \caption{Poloidal mode number scan ($m=6\dots10$) of the $n=4$ shear Alfv\'{e}n wave in toroidal Cyclone-like geometry for $\beta_e=3.3\cdot 10^{-3}$. The dotted line is the analytic mode frequency in cylindrical geometry.}
 \label{fig:alfven_dispersion_toroidal}
\end{figure}

Similar tests in toroidal geometry have been performed in a slightly modified version of the standard cyclone geometry, with $R_0=1.7$, $a/R_0=0.358$, $B_0=1.9$, constant $q=2$, and $T_0=2$ keV.
The density is varied between $1.875 \cdot 10^{19}\,\mathrm{m}^{-3}$ and $6 \cdot 10^{20}\,\mathrm{m}^{-3}$ to achieve values of $\beta_e$ between $0.4$\% and $13.4$\%.
The time step is $\Delta t \approx 5\cdot 10^{-2} \tau_A$ and the total simulation time is $\approx 40 \tau_A$.
Figure \ref{fig:alfven_dispersion_toroidal} a) shows a poloidal wave number scan ($m=6\dots10$) of the frequency of the $n=4$ shear Alfv\'{e}n wave.
The numerical frequencies agree very well with the (approximate) analytical result $\omega \propto (2\pi/L_\parallel) (n-m/q) v_A$, where $L_\parallel$ is the parallel connection length for one poloidal circuit.
The deviations are caused by the variation of the field line pitch along magnetic field lines.
We find that $\omega \tau_A$ is independent of $\beta_e$ as expected because only the density $n_0$ was varied in this test.
%

%
Since the kinetic-fluid hybrid approach is especially useful for the simulation of low-n tearing modes, we benchmarked the $(m,n)=(2,1)$ tearing mode in cylindrical and toroidal geometry against the GEM code \cite{chen_2015} and M3D-K \cite{cai_2012}, respectively.
We do not consider the effect of kinetic ions in this benchmark.
The only term added to the electron fluid equations compared to the terms kept in the shear Alfv\'{e}n case is the kink drive $\deltab \cdot \nabla [j_0/(eB)]$ in Eq. \eqref{eq:continuity_eq} to be consistent with GEM's eigenvalue solver \cite{chen_2015}.

\begin{figure}
 \centering
 \includegraphics[bb=0 0 208.461945 300.292816]{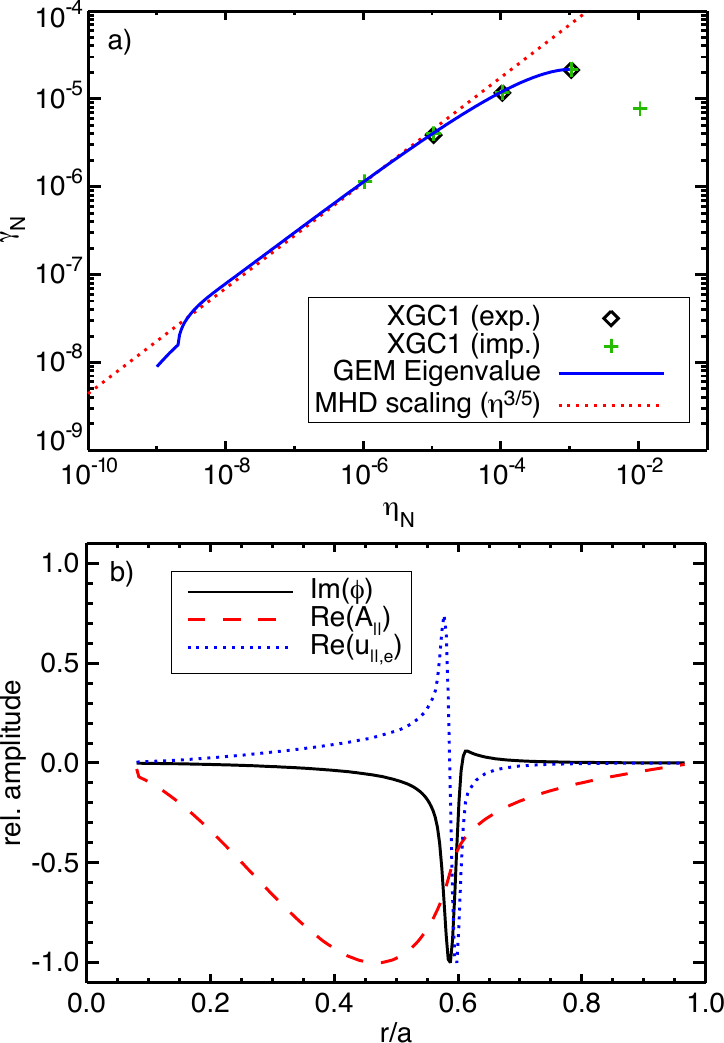}
 \caption{a) Growth rate of the $(2,1)$ tearing mode in cylindrical geometry. The solid line shows the result of GEM's tearing mode eigenvalue solver in the MHD approximation for the ion polarization density (see Ref. \onlinecite{chen_2015}).
 The results from the implicit and explicit time integrator agree very well.
 b) Mode structure of the $m=2$ mode with $\eta_N=10^{-5}$. The amplitude of each quantity is normalized to its respective maximum.
 The mode structure compares well to Fig. 8 in Ref. \onlinecite{chen_2015}, although the tearing layer is shifted to slightly higher $r/a$.}
 \label{fig:tearing_gamma_cyl}
\end{figure}

For the benchmark against the GEM code, we use the case described in Ref. \onlinecite{chen_2015}: concentric, circular flux-surfaces in cylindrical geometry, $R_0=1.7$ m, $a=0.425$ m ($R_0/a=4$), $B_0=1.906$ T, $q = 1.5 [1+(r/a)^2]$, $Z=1$, $m_i/m_p=2.5$, and constant density $n_0=3.886 \cdot 10^{20}\,\mathrm{m}^{-3}$.
Since the electron fluid equations used for this benchmark have no temperature dependence, we can use a constant temperature profile $T_{0,e}=45.63$ eV, which yields the same on-axis $\beta_e$ of $4 \cdot 10^{-3}$ and relative domain size $a/\rho_i \approx 740$ as in Ref. \onlinecite{chen_2015}.
The resonant surface for the $(2,1)$ tearing mode is at $(r/a)_{c} \approx 0.577$ corresponding to the normalized poloidal magnetic flux $\psi_{N,c}=0.411$.
In order to be able to resolve the resonance layer of the $(2,1)$ tearing mode also at low resistivity, the radial resolution of our computational mesh varies between $0.5$ mm around the resonant surface to a maximum of $8$ mm far from the resonant surface.
The relations between the normalized resistivity $\eta_N$ and growth rate $\gamma_N$ used in Ref. \onlinecite{chen_2015} and the corresponding values $\eta_e$ and $\gamma$ in SI-units are $\eta_N = (e n_0/B_0) \eta_e$ and $\gamma_N = m_p/(e B_0) \gamma$, where $m_p$ is the proton mass.
The results of a resistivity scan of the growth rate of the $(2,1)$ tearing mode in this geometry is shown in Fig. \ref{fig:tearing_gamma_cyl}.
The growth rates evaluated with XGC1 show excellent agreement with the growth rates computed with GEM's eigenvalue solver that uses the MHD approximation for the ion polarization density (Fig. 3 in Ref. \onlinecite{chen_2015}).
We did not include an XGC1 data point for $\eta_N=10^{-7}$ because of the very strict resolution requirements of about $2.5 \cdot 10^{-4}$ m or less for this low resistivity.
Using the Crank-Nicolson method, the implicit time integrator could speed up these simulation by a factor of more than 10.
For $\eta_N=10^{-6}$ a time step of $\Delta t = 2.7 \tau_A$ could be used.

\begin{figure}
 \centering
 \includegraphics[bb=0 0 197.521576 512.728577,scale=0.9]{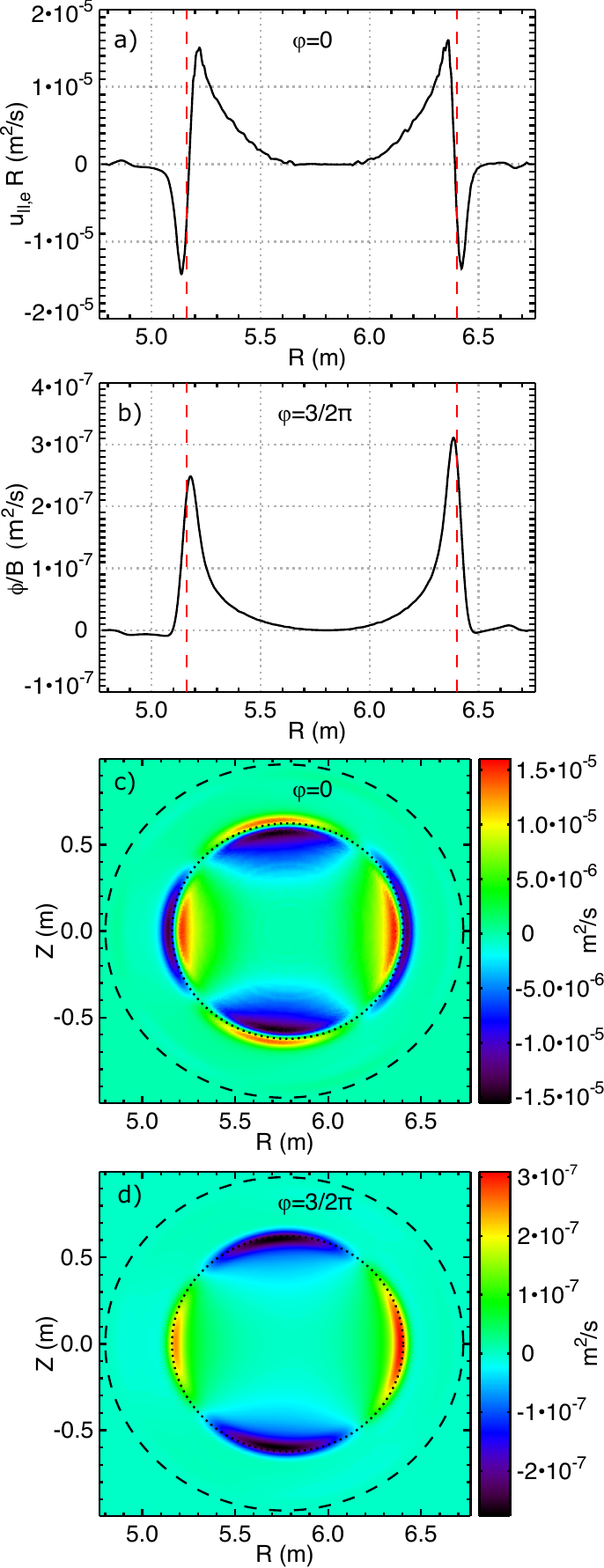}
 \caption{Mode structure of the $(2,1)$ tearing mode in toroidal geometry, which compares well to Fig. 1 in Ref. \onlinecite{cai_2012}.
          a) $R u_{\parallel,e}$ (equivalent to the perturbed current) at $\varphi=0$, and
          b) $\phi/B$ (velocity stream function) at $\varphi=3\pi/2$ plotted along the midplane.
          The dashed lines indicate the location of the $\psi_N=0.5$ surface.
          c) $R u_{\parallel,e}$ at $\varphi=0$, and
          d) $\phi/B$ at $\varphi=3\pi/2$ in the $(R,Z)$ plane.
          The dotted circle is the $\psi_N=0.5$ surface.}
 \label{fig:tearing_tor_mode_structure}
\end{figure}

For the benchmark against M3D-K in toroidal geometry, we use a Grad-Shafranov equilibrium generated with the FLOW code \cite{guazzotto_flow_code} with a fixed circular boundary, $R_0=5.76$ m, $a=1$ m, $B_0=1$ T, $q = 1.5 + 2 \psi_N^2$, $m_i/m_p=2.5$, and constant $n_0=10^{20}\, \mathrm{m}^{-3}$ and $T_{0,e}=100$ eV, so that $\beta_e = 4 \cdot 10^{-3}$ and $\beta_e m_i/m_e = 18.4$.
The resonant surface of the $(2,1)$ tearing mode is located at $\psi_N=0.5$.
The radial resolution of the computational mesh is  $1.5$ mm between approximately $\psi_N=0.4$ and $\psi_N=0.6$ and up to $1.2$ cm away from the tearing layer.
For the normalized resistivity of $\eta_{M3D}=10^{-4}$ used in Ref. \onlinecite{cai_2012} and the corresponding normalization relations, we obtain a resistivity in SI units of $\eta_e = \mu_0 (a^2/\tau_A) \eta_{M3D} = 3.01 \cdot 10^{-4}$.
The XGC1 growth rate calculation used a time step of $\Delta t = 7 \cdot 10^{-3} \tau_A$ and ran for a total time of approximately $350\, \tau_A$.
Figures \ref{fig:tearing_tor_mode_structure} a)-d) show the mode structure of the growing $(2,1)$ mode, which exhibit the usual tearing structure.
For comparison with Ref. \onlinecite{cai_2012}, we use reduced MHD quantities, the perturbed current $R u_{\parallel,e}$, and the velocity stream function $\phi/B$.
The growth rate we obtain from the XGC1 calculation is $\gamma = 1.12 \cdot 10^{-2} \tau_A^{-1}$ and compares well to Ref. \onlinecite{cai_2012}.
The relative difference between the XGC1 and the M3D-K result $6$\%.

%
%
In order to add gyrokinetic ion effects to electromagnetic fluid/MHD instabilities, the gyrokinetic edge turbulence code XGC1 has been modified by replacing the kinetic electrons by massless drift-fluid electrons \cite{parker_2000,chen_2001}. 
Explicit and implicit time integration methods have been implemented and tested.
We verified shear Alfv\'{e}n wave physics against the analytical solution and benchmarked the massless fluid model for resistive tearing modes against the codes GEM and M3D-K.
The hybrid model in XGC will be further developed into a total-f code with the aim of studying the onset of edge localized modes across the magnetic separatrix surface.
Verification of the kinetic version of peeling-ballooning modes, and kinetic ballooning modes will be reported in a subsequent paper.\\

%
The authors would like to thank Guoyong Fu, Stephen Abbott, and Eduardo D'Azevedo for fruitful discussions.
Support for this work was provided through the Scientific Discovery through Advanced Computing (SciDAC) program funded by the U.S. Department of Energy Office of Advanced Scientific Computing Research and the Office of Fusion Energy Sciences.
Notice: This manuscript has been authored by Princeton University under Contract No. DE-AC02-09CH11466 with the U.S. Department of Energy. The publisher, by accepting the article for publication, acknowledges that the United States Government retains a non- exclusive, paid-up, irrevocable, world-wide license to publish or reproduce the published form of this manuscript, or allow others to do so, for United States Government purposes.


%


\newpage
\renewcommand{\refname}{}
\bibliography{references}


\end{document}